# Study of the oxidation process of crystalline powder of $In_2S_3$ and thin films obtained by Dr Blade method

M. Hamici [1], S. Guessoum [1,2], L. Vaillant [3], Y. Gagou [3], and P. Saint-Grégoire [5]

(1) DAC laboratory, University Sétif-1, Algeria

(2) University A. Mira of Bejaia, Algeria

(3) Photovoltaic Research Laboratory, Institute of Materials Science and Technology - Physics Faculty, University of Havana, 10400 Havana, Cuba

(4) LPMC, Université de Picardie Jules Verne, 80039 Amiens Cedex 01

(5) University of Toulon, Campus of La Garde – La Valette, 83130 La Garde, France

**Corresponding author: Pierre Saint-Grégoire, pstgregoire@gmail.com**

Abstract :

Good quality $In_2S_3$ films were grown by Dr Blade method from a powder synthesized in a chemical bath, and oxidized to obtain $In_2O_3$ films and films of intermediate composition. The oxidation process and kinetics are studied by means of thermogravimetric analysis, which shows complex phenomena. The study is completed by scanning electron microscopy and EDX measurements. X-ray diffraction and UV-visible spectroscopy results are also shown. The oxidation process begins significantly above 420 C, and it appears that at least one intermediate crystal phase takes place in the solid solution, suggesting that the replacement of sulfur by oxygen atoms occurs at different temperatures in the different crystallographic sites. The obtained samples show a band gap varying continuously between 1.94 eV and 3.72 eV.

## I. Introduction

$In_2S_3$ is an interesting material that has been studied in the last years mainly in reason of its potential interest in solar cells as a CdS buffer layer substitute for fabricating Cd-free CIGS solar cells [1, 2, 3, 4, 5]. Most of $In_2S_3$ films of the crystalline β form [6] reported in literature are obtained by Chemical Bath Deposition (CBD) [7, 8], physical vapor deposition (PVD) [1], flash evaporation [9], or spray pyrolysis [10]. More recently atomic layer deposition (ALD)[11], pulsed laser deposition (PLD) [12], and magnetron sputtering [13] have also been used. $In_2O_3$ is on the other hand in several aspects an awesome material existing mainly in a cubic symmetry [14] when the synthesis is not realized at high temperatures and pressures or in non-equilibrium growth conditions. This material is well known to be used in particular in transparent electrodes when doped with Sn for liquid crystals displays, photovoltaic devices, and LEDs [15]. Both materials, $In_2S_3$ and $In_2O_3$, are semiconductors, with different band gaps around 2 - 3 eV [16, 17].

In the literature the role of oxygen introduced within $In_2S_3$ films was described for films obtained by CBD [16], spray pyrolysis [17] and PVD [16]. Besides the properties of $In_2O_3$ obtained by oxidation of $In_2S_3$ synthesized by spray pyrolysis were investigated in [20]. Some of us showed using $In_2S_3$ films grown by CBD that the oxidation process depends on the film microstructure [19]: the temperature region in which the oxidation occurs is considerably lower when the size of crystals in deposited films is nanometric, and the temperature of annealing in air atmosphere in this case is observed to determine the composition of the solid solution.

A potential interest of intermediate compositions between $In_2S_3$ and $In_2O_3$ is that the band gap varies as a function of composition [16] and can be tailored to intermediate values between the band gap value for $In_2S_3$ and that for $In_2O_3$, which should influence the behaviour of solar cells or other devices built with the corresponding buffer layer. However on the other hand, it should be also considered that nanostructured thin films are fragile with respect to heating or also to chemical attacks. Attempts to deposit the absorbers p-type semiconducting materials $Cu_2SnS_3$ (CTS) and $Cu_2ZnSnS_4$ (CZTS) on $In_2S_3$ thin films by chemical methods, with an interpenetration of materials, is difficult because of chemical instability of the $In_2S_3$ film in the corresponding baths [20].

This situation motivated us to investigate the possibility to obtain $In_2S_3$ films by Dr Blade technique, and to study their oxidation. We thus carefully studied the behaviour of such films and constitutive crystalline powders, that have grains with a micrometric size as evidenced in the SEM, to be compared with results obtained with nanometric powders resulting from thin films as studied in [21].

In the following we present first the films obtained by Dr Blade method, and then results of TGA that show mass losses in nitrogen atmosphere, resulting from the elimination of solvents ; we thereafter show mass losses in air atmosphere, accompanying the substitution of sulphur by oxygen in the structure. In TGA experiments, both ramp modes and isothermal modes were used. The mechanism involving oxidation is confirmed by analyses performed by EDX and X-ray diffraction (XRD). The variation of band gap of the obtained materials of different compositions is deduced from UV-visible spectrometry data.

## II. Experimental methods

The method for sample synthesis was derived from that given by Sandoval et al for deposition of thin films by chemical bath deposition [8] : the preparation starts with the dissolution of 0.1 M $InCl_3$ in 10 ml of water, to which 20 ml of 0.5 M acetic acid and finally 20 ml of 1 M thioacetamide are added, the volume being finally completed till 100 ml with distilled water.

The synthesis of $In_2S_3$ occurs according to the chemical process:

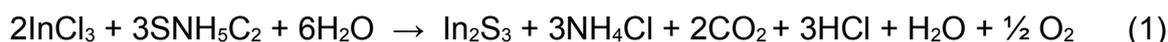
$$2InCl_3 + 3SNH_5C_2 + 6H_2O \rightarrow In_2S_3 + 3NH_4Cl + 2CO_2 + 3HCl + H_2O + \tfrac{1}{2} O_2 \quad (1)$$

The resulting mixture is then heated till 70 C and continuously stirred. Powders were collected by filtering the final preparation, and dried in air. The deposition of layers by Dr Blade method was inspired from the preparation described for $TiO_2$ in [23]. We slightly varied the proportions, and remarked that adding between one and three droplets of ethylene glycol in the preparation reduces the occurrence of cracks in the layers, and even leads to samples completely exempt of any fissure.

Thermogravimetric measurements were performed using an apparatus 2950 TGA (TA Instrument), in two modes : ramp mode, with a constant temperature rate (generally between 10 C/min and 20 C/min), and isothermal mode (versus time) in nitrogen atmosphere and in air atmosphere. Scanning electron microscopy (SEM) was performed by means of Hitachi S4500 and S4800 instruments, respectively for image and EDX studies. In the EDX study of powder composition, we gently compressed the sample on a glass plate, in order to obtain a flat surface to perform measurement. UV-visible spectrometry was performed using a Jasco V650 spectrophotometer in standard conditions. For XRD measurements we used a powder diffractometer Philips X-PERT PRO II working with CuKα1 and CuKα2 radiation, in classical Bragg Brentano geometry. Data were analyzed using the Fullprof software [24].

## III. Experimental results and discussion

### III.1. Morphology of the samples

The resulting material (Figure 1) shows grains with sizes distributed around an average value around 0.6 μm, decorated by nanocrystals similar to those deposited by CBD in thin films, that occupy a negligible volume in the sample.

Layers obtained by Dr Blade method are homogeneous, and exempt of cracks when few drops of ethylene glycol are added in the preparation. Figure 2 shows a planar view obtained by scanning electron microscopy, that indicates that over a length as large as 2 mm no crack is observed. The insets show details of the same sample. The cross section view is presented in the left inset, showing from the bottom to the top, the glass plate, the thin film of conducting oxide, and the In$_2$S$_3$ layer. The right inset presents a planar view with a higher magnification.

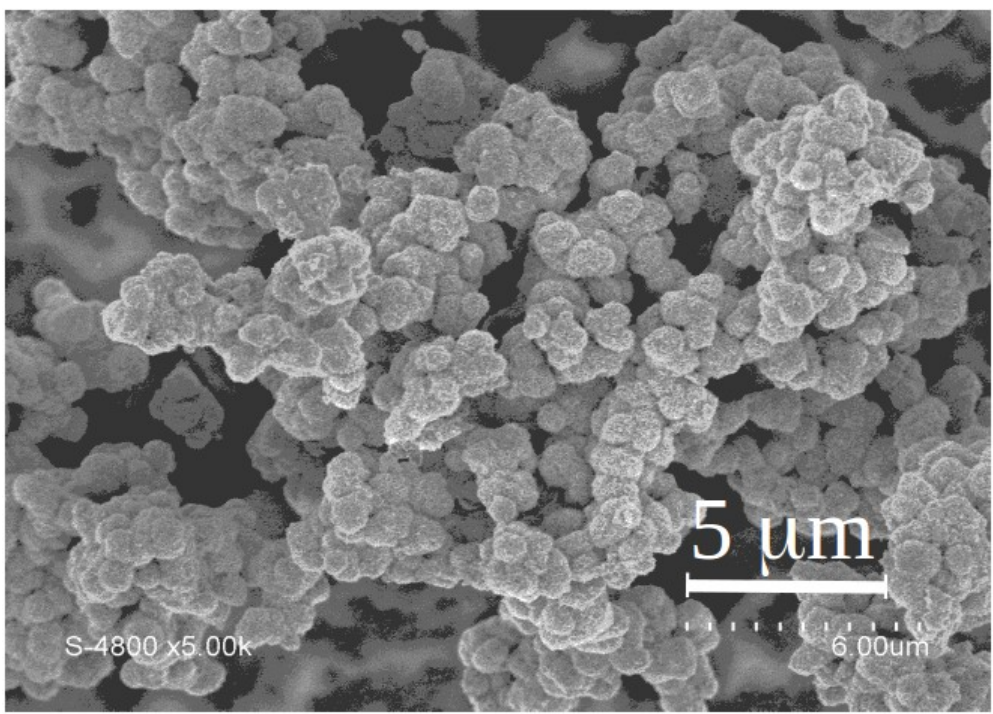

**Figure 1: SEM picture showing the microstructure of In$_2$S$_3$ powder**

The sample is the as-grown powder ; the grains appear to have an average diameter around 600 nm.

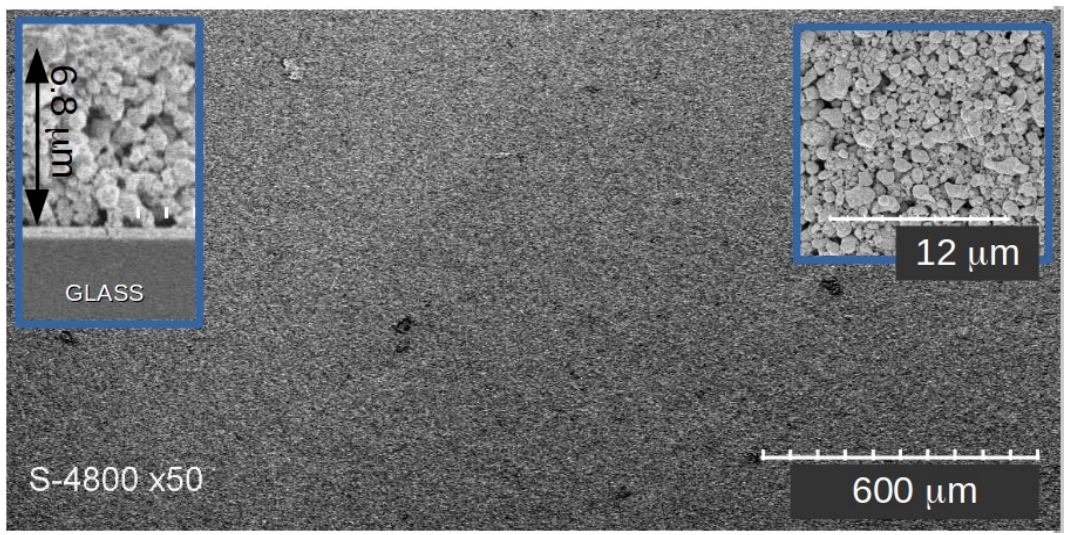

**Figure 2: SEM picture of an In$_2$S$_3$ layer deposited on a glass plate**

Note that this sample is homogeneous and exempt of any crack ; the right inset shows a region of the sample at a higher magnification, the left inset presents a cross-section view of the same sample. In this last inset, the upper layer is In$_2$S$_3$ and the thin intermediate dense layer between it and the glass is the conducting oxide (here FTO).

## III.2. Results of TG measurements

Figure 3 presents the TG results in N$_2$ atmosphere on the first heating and on the next temperature run. We observe sharp mass losses attributed to evaporation of solvents during the first heating (run AB) ; in further measurements the mass appears indeed to remain stable both on cooling (run BC) and further heatings (run CD and further runs not shown in the figure) in this inert gas during the ramps.

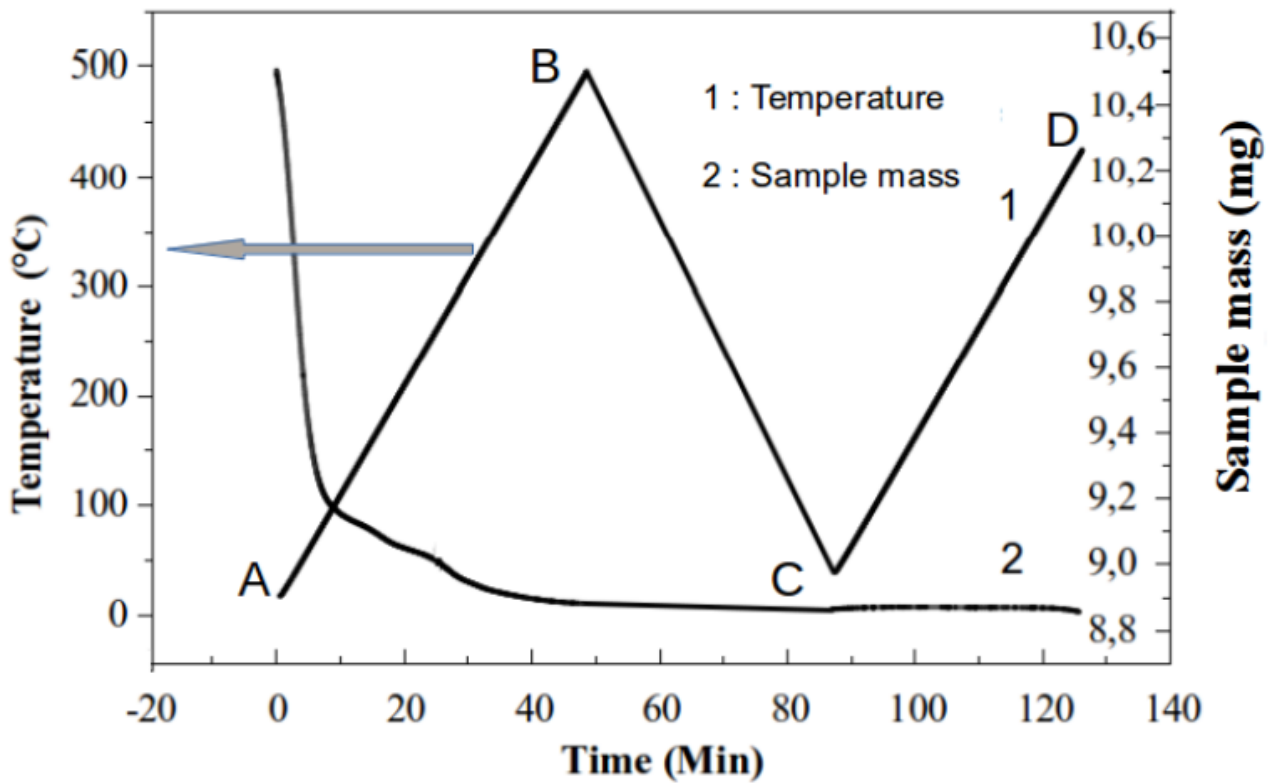

**Figure 3: TG measurements in a N$_2$ atmosphere**

The mass loss occurs during the first heating (curve AB) of the as-grown powder. On further cooling (curve BC) and heating (curve CD) the mass remain stable.

When the heating occurs in air there occur mass losses (Figure 4) attributed to the substitution of sulphur atoms by oxygen atoms (we shall see later on in the present paper that this is confirmed by complementary studies). Figure 4 shows also the derivative of mass loss, that better exhibits the temperatures at which the rate of oxidation presents an anomaly. We thus see in the derivative a main peak pointing around 520 C, accompanied by a shoulder around 480 C ; the process begins around 420 C and is strong till approximately 550 C. Above this last temperature we still observe a mass loss but with a much smaller slope, showing that the process of oxidation is still incomplete and continues to develop. This feature could be due to a kinetic process still continuing above 550 C.

The results shown in Figure 4 suggest that the oxidation of the compound occurs till 480 C as a perturbation of the $In_2S_3 – \beta$ structure (tetragonal), whereas the process starting from 550 C (foot of the main peak in the derivative) could be associated with the progressive oxidation continuing in the $In_2O_3$ structure (cubic).

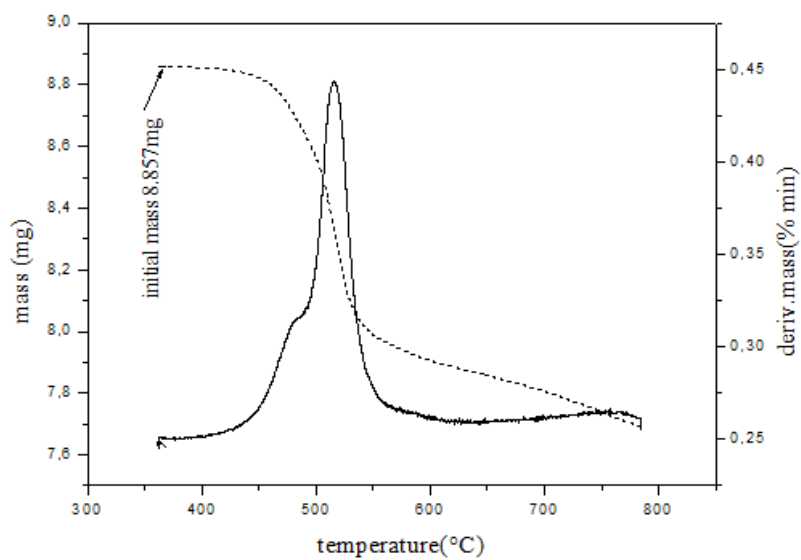

**Figure 4: TG measurements and derivative of mass loss in an air atmosphere**

The mass loss is represented by the dashed line. The derivative of mass (full line) presents a main peak and a shoulder, which evidence two particular temperatures, around which a specific process occurs. This suggests an intermediate step between them, in the oxidation process.

In order to determine if kinetic processes may be observed, and to differentiate the kinetic part of what depends only on temperature, we performed fast changes of temperatures followed by isothermal measurements, starting with the original sample. The results are shown in Figure 5. The first temperature increase is performed at a scan rate of 100 C/min (curve AB) till 470 C in an atmosphere of nitrogen. We observe a mass decrease due to departure of residual products expected from equation (1), similar to that shown in Figure 3. Thereafter we immediately changed the gas to air and performed an isotherm during 30 min (curve BC), then another fast temperature change to 485 C during (15 min) (above point C) and an isotherm of 10 min (curve DE). Finally we performed a last fast scan till 750 C (above point E).

These data evidence indeed kinetic phenomena: at the beginning of the first isotherm in air, we observe a mass decrease due to oxidation and departure of sulphur from the crystalline structure. Surprisingly, the mass presents however a local minimum and then increases again with time, which we interpret as a possible reversed mechanism of sulphurization ; the possible mechanism could be that when oxygen atoms enter in the structure (at the beginning of the isotherm in air), there occurs a release of sulphur atoms and as a consequence the sample is in a S-rich atmosphere which leads to its partial re-sulphurization.

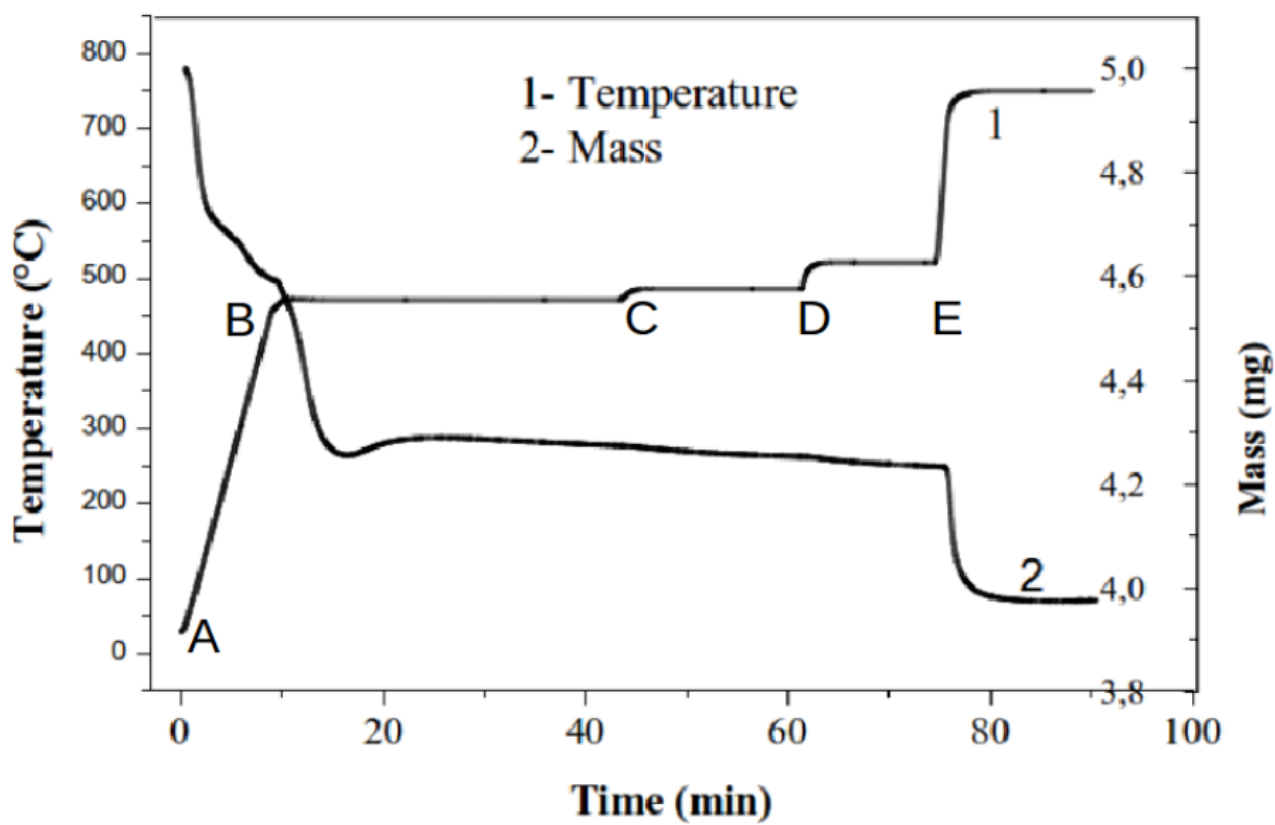

**Figure 5: Fast heatings followed by isothermal measurements of mass losses**
During the fast heating at 100 C/min (curve AB), the sample is in a $N_2$ atmosphere, thereafter the gas is changed to air and we observe mass changes due to oxidation, as confirmed by complementary studies. The change of temperature at point E leads to a mass decrease. Similar phenomena are observed at points C and D, but are too weak to be observed clearly in the present curve (see details in Fig. 6)

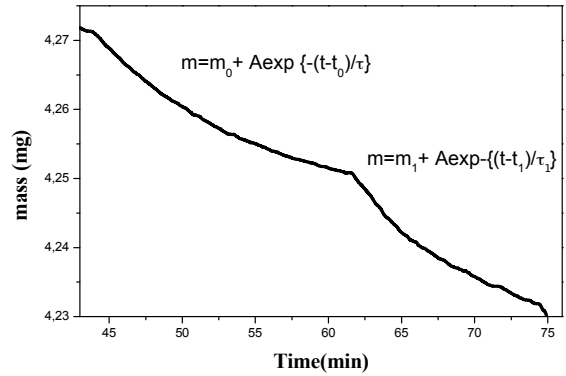

**Figure 6: mass versus time during TG isothermal measurements around points C (left) and D of Figure 5.**

To better understand the processes, we performed various quick heatings followed by temperature stabilization and isothermal measurements. Thus, after 30 min at 470 C, we changed quickly the temperature to 485 C (point C in Figure 4) and observed a change of regime in the decrease of mass, showing that oxidation is enhanced by the temperature increase; this type of behaviour is also observed after this isotherm, by heating quickly to 520 C (point D). A behaviour of the same type, but much stronger, is observed when heating quickly from 520 C till 750 C. In all cases a clear exponential mass decay is observed. Figure 6 shows a magnification of the curve of Figure 5 around point D ; the anomalous part of mass change with time may be fitted by a law of exponential decrease:

$$m(t,T) = m_0(T) + A(T)\exp{-(t-t_0)/\tau(T)} \quad (2)$$

where $t$ is the time, $T$ the temperature, $\tau(T)$ the characteristic time of the reaction. $m_0$ is the mass reached asymptotically at a given temperature, determining the composition. In this expression $\tau$ , $A$, and $m_0$ are indeed determined to be dependent on the annealing temperature in air atmosphere; is determined to vary from 8.3 min at 476 C to 2.2 min at 746 C. These values are comparable to those determined in the nanometric power but at much lower temperatures [21].

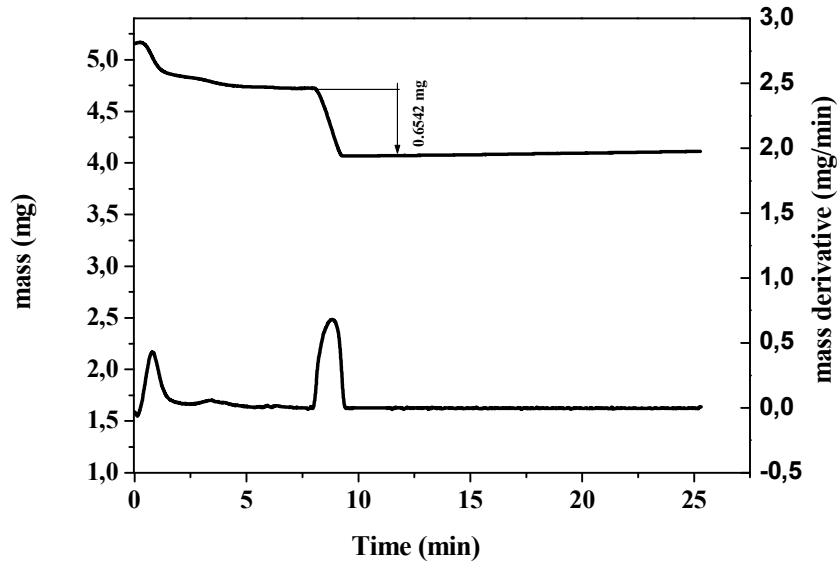

**Figure 7: TG measurements exhibiting the fast oxidation at high temperature**

The upper curve represents the mass loss versus time obtained on heating first in $N_2$ atmosphere, and then at high temperature (800 C in the present case) when changing the atmosphere to air. The downer curve represents the derivative, that better exhibits the phenomena.

The results shown in Figures 4 and 5 strongly suggest that the process is quasi-complete only above 750 C. It is why we performed a last study consisting in a fast heating (100 C/min) in nitrogen till 800 C, followed by a change of gas to air at high temperature. The result is shown in Figure 7; it shows that the process occurring as soon as the gas is changed to air, is fast : the oxidation occurs in a time interval of about 1 min, and then the mass is stable.

Considering that the mass of the sample assumed pure $In_2S_3$ is 4.5 mg (mass just before the entrance of air) leads to conclude that this sample contained 0,0138 mol (for $In_2S_3$ we have 325,82 g/mol). This quantity of moles for $In_2O_3$ has a mass of 3,83 mg (the molar mass of this last compound is 277,64 g/mol), corresponding to a mass difference of 0.67 mg with respect to the starting mass, to be compared with the experimental mass loss of 0.6542 mg, which certainly hinders at the presence of residual sulphur atoms in the compound treated at 800 C. The complete oxidation of $In_2S_3$ into $In_2O_3$, taking into account the differences of molar masses, should lead to a mass decrease of ~14.78%, whereas we never measured such a high mass decrease, the highest obtained value is ~13.7%, that indicates that the final product still contains sulphur atoms.

## III.3. Scanning electron microscopy and EDX measurements

To check that it is indeed an oxidation process that occurs, we performed studies on samples treated at various temperatures corresponding to interesting points in the TG curve obtained in air (Fig.4): after a treatment in nitrogen atmosphere of all samples at 300 C to remove residuals from the sample, samples were respectively heated till various temperatures between 300 C and 650 C. Before observation and measurements they were gently compressed against a small plate of glass, to reduce the relief in order to obtain samples morphologically similar, which allows semi-quantitative composition determinations and comparative studies of composition between samples.

Results obtained on samples treated at 300 C, 520 C, and 650 C are shown in Figure 8. We see that the morphologies are rather identical, which is a favorable situation to compare EDX results.

In the sample heated till 300 C, the analyses performed in different parts of the sample give average results of 7.5 at% of oxygen, 55.25 at % of sulphur, and 37.25 at% of indium. As we know from the synthesis that this original sample should have the chemical composition $In_2S_3$, the presence of oxygen in the spectra is attributed to the substrate (FTO conducting glass, namely $SnO_2$ doped with F).

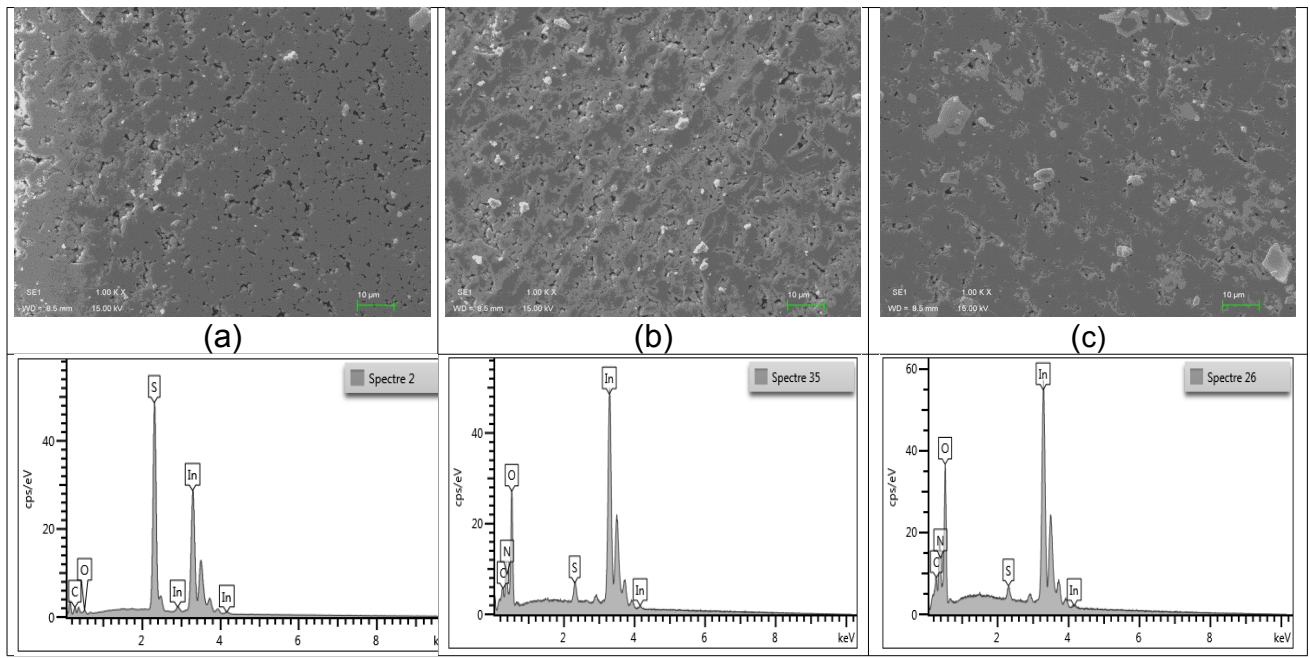

**Figure 8: SEM pictures and results of EDX analyses**
(a) sample heated at 300 C in air  (b) sample heated at 520 C  (c) sample heated at 650 C

In the sample heated till 520 C, there appears that oxygen atoms have indeed more massively penetrated into the structure whereas the peak of sulphur has fallen down but is still present. Average results over different regions give rather satisfactory results since the ratio between In and S + O is 1.45, close to the value 1.5 expected from the chemical formula and an O/S ratio of about 88 at% O, and 12 at% S corresponding to the chemical formula $In_2S_{3(1-x)}O_{3x}$ with x~ 0,88, the sample having the exact composition $In_{2,07}S_{0.36}O_{2.64}$.

In the sample heated till 650 C we observe that the oxygen atoms concentration in the structure has increased significantly as shown by the higher O peak and the smaller S peak in the spectrum (Figure 8-c): we deduce now a O/S ratio with 97,07 at% oxygen and 2,93 at% S, corresponding to the x value x~0.9707 and the chemical formula $In_2S_{0.088}O_{2.91}$. For the sample heated till 800 C, measurements confirm that there are only traces of S, and the oxygen peak is much predominant.

It is interesting to compare these results to those expected from the data of Figure 4, attributing the mass loss to oxidation and to the molar mass difference between sulphur and oxygen. This is shown in Figure 9. We observe that there is a systematic difference, the values obtained by EDX being above the expected values from thermogravimetric data. This may be explained by two factors: the first one concerns the kinetic phenomena; TGA data represented in Figure 9 were indeed obtained in the ramp mode, at 10 C/min and do not correspond to equilibrium values since the oxidation phenomenon occurs with characteristic times of several minutes depending on temperature. One thus has to consider that the equilibrium curve should be obtained by distorting the curve of Figure 9 in such a way as leaving unchanged the left part but by modifying the right part in order that it terminates at higher values closer to 1. A second origin of uncertainty lies in the EDX measurements that are performed on samples having a certain roughness (compacted powders); moreover the substrate makes obviously the measurements difficult in reason of the presence of oxygen in the substrate (glass).
If the exact composition cannot be determined with accuracy, we may nevertheless deduce that a partial oxidation begins around 420 C whereas around 800 C the oxidation is quasi complete.

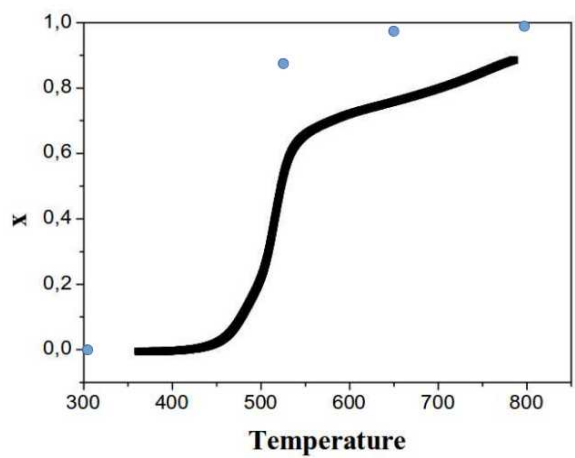

**Figure 9: Comparison between expected composition parameter x and measured by EDX**
The full curve represents the expected composition parameter from TGA measurements whereas dots represents results of EDX.

During such a process of oxidation, there could happen that oxygen penetration is inhomogeneous, it is why it is important to pay attention to the spatial distribution of In, O and S elements. The map of concentrations obtained in the microscope indicates however that the composition is homogeneous in the sample.

**III.4. X-ray diffraction results**

Powders originating from the same sample and treated at various temperatures in air atmosphere were studied by means of XRD. A special attention has been paid to samples corresponding to interesting temperatures in TG results.

The results are shown in Figure 10, for samples heated till 485 C. We see that all curves present globally the same appearance, except that of the sample treated at 485 C, that presents a peak absent in other curves.

The sample treated at 300 C presents a diffractogram corresponding well with the pure $In_2S_3$ in its β form, as shown by the Rietveld refinement result shown in the inset of Figure 10. In this inset, we present experiment data, calculated XRD pattern using Rietveld method and their difference highlighting the good quality of obtained results. Small vertical lines in the intermediate position are the theoretical positions of the main Bragg reflections. We see that for the curves labeled according to the treatment temperature (300, 450, and 475 C) the main peaks are approximately at the same positions whereas changes are mainly observed on the intensity of Bragg reflections. This shows that, as assumed precedently, the structure remains the same, and oxidation occurs by replacement of sulphur atoms within the $In_2S_3$-β structure.

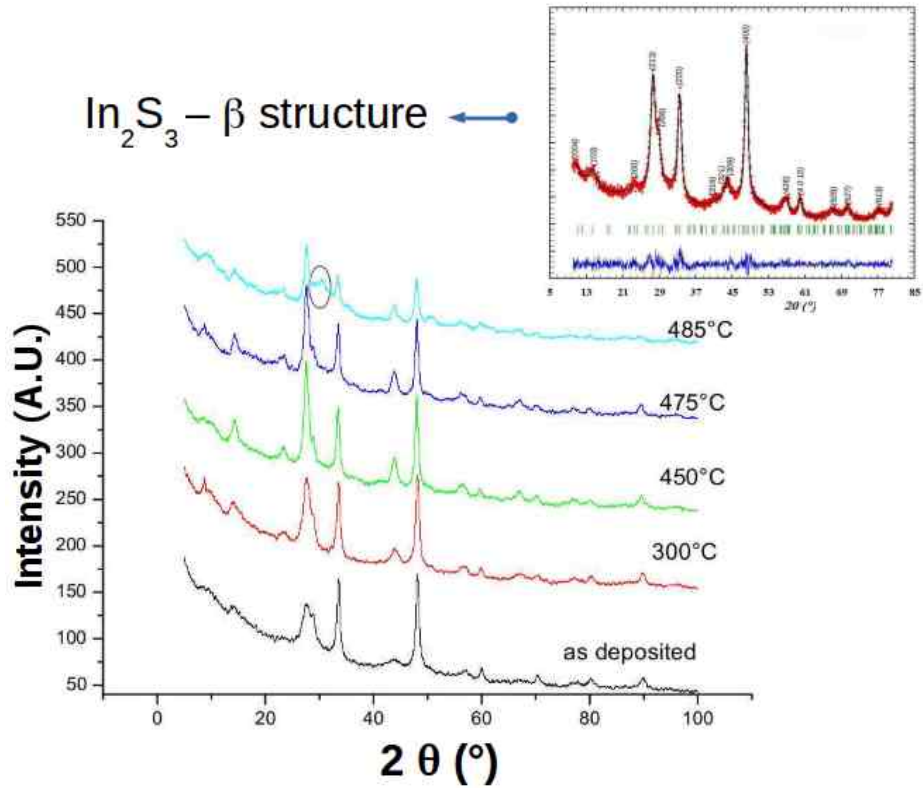

**Figure 10: Diffraction patterns of samples treated in air atmosphere at different temperatures**

The inset is the result of the Rietveld refinement of the curve corresponding to the sample treated at 300 C, which confirms the $In_2S_3$-β-structure. The upper curve of the inset represents the experimental points, and the result of the calculated spectrum by the Rietveld method. The curve below represents the difference between experimental data and the calculated profile, whereas the small vertical lines show the positions of the main Bragg reflections.

The other curves are remarkably globally similar; the changes are mainly on the intensity of reflections, which shows that the global crystal structure is unchanged between them. Only the pattern obtained on the sample treated at 485 C shows a peak absent in patterns of other samples.

Diffraction patterns obtained on samples treated at a higher temperature are different. We see in Figure 11a and b, the diffraction patterns obtained on samples treated between 485 C and 1000 C.

Figure 11a presents the diffraction patterns obtained at 485 and 520 C, namely in the region between main peaks in the TG signal (see Figure 4). The curve of the sample treated at 485 C still resembles the $In_2S_3$ type diffraction pattern, but we notice the clear presence of a new diffraction line located just above $2\theta \sim 30°$. The compounds treated in air at 520 C and 550 C present a different diffraction pattern, but shows the same intense line.

As shown in diffraction studies of other samples, treated at temperatures above 550 C, this line remains while we observe the disappearance of several peaks and the increase of others. The pattern observed on the sample treated at 1000 C coincides with that of pure $In_2O_3$, and it appears that the most intense reflection in preceding patterns corresponds with the (222) reflection in the diffraction pattern of $In_2O_3$ as shown in Figure 12 (see also Table 1 resuming the results of Rietveld refinement).

These data show that there are indeed 3 different regions that may be distinguished: in the first one, containing samples treated between 300 C and 475 C, we have an $In_2S_3$ type crystal structure, in the third one we have compounds treated above 550 C, with an $In_2O_3$-type structure, and in between we have the second region with a still undetermined crystal structure. Thus, diffraction results are in good agreement with conclusions that may be obtained from thermogravimetric analysis.

To attempt to understand these results, we may consider that the sulphur atoms occupy different crystallographic sites in the β-structure, namely sites with different environments. It is then natural to assume that the replacement of S atoms by O atoms, occurs in different sites with different probabilities, thus giving rise to a new, intermediate phase, when some sites are all occupied by oxygen atoms. Complementary studies and a detailed analysis of diffraction data by the Rietveld method are necessary to check this hypothesis and determine the intermediate phase.

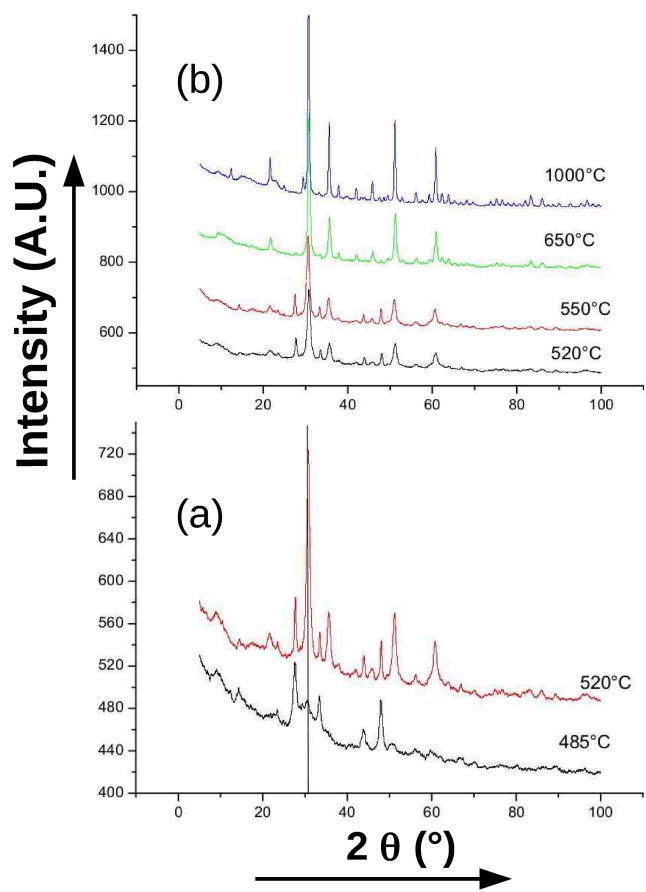

**Figure 11: Diffraction**

**patterns obtained on samples treated at different temperatures above 485 C**

In (a) we observe the patterns obtained for samples treated at 485 C and 520 C, which shows the strong increase of a line located around 2θ= 30° and a pattern that distinguishes itself from others.

In (b) are shown the spectra obtained at higher temperatures, where we see the progressive transformation to $In_2O_3$.

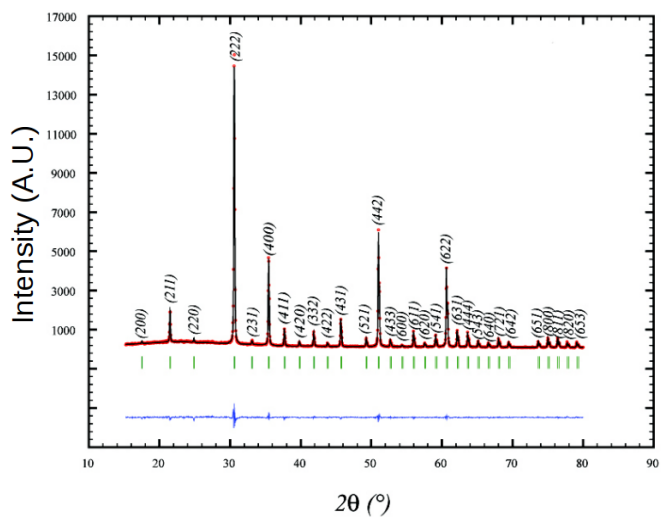

**Figure 12: Diffraction pattern of the sample treated at 800 C, and results of Rietveld analysis**

The upper curve shows the experimental points and the calculated profile, the quality of which may be estimated through the curve showing the difference between them (below). Vertical lines show the position of Bragg reflections.

Finally, the diffraction pattern of the sample treated at 800 C is shown in Figure 12 that presents also the result of Rietveld refinement confirming that the resulting phase corresponds well with $In_2O_3$.

**Table 1: Results of Rietveld refinement on two samples**

| *Samples* | Thermal treatment in air at 800 C | Thermal treatment in air at 300 C |
|---|---|---|
| *Space Groups* | I a -3 (N° 206) | I 41/a m d (N° 141). |
| *Crystal systems* | cubic | tetragonal |
| **Source wavelength** | $\lambda CuK\alpha_1 = 1.54051$  $\lambda CuK\alpha_2 = 1.54433$ | $\lambda CuK\alpha_1 = 1.54056$ |
| **Lattice parameters** | | |
| **a(Å)** | 10.1204 | 7.5866 |
| **b(Å)** | 10.1204 | 7.5866 |
| **c(Å)** | 10.1204 | 31.9703 |
| **V(Å³)** | 1036.5524 | 1840.1008 |
| **Reliability factors** | | |
| R<sub>F</sub> factor | 0.918 | 4.95 |
| Bragg R factor | 0.875 | 0.144 |
| Rwp | 12.6 | 20.4 |
| Chi² | 1.719 | 1.34 |

### III.5. UV –Vis measurements

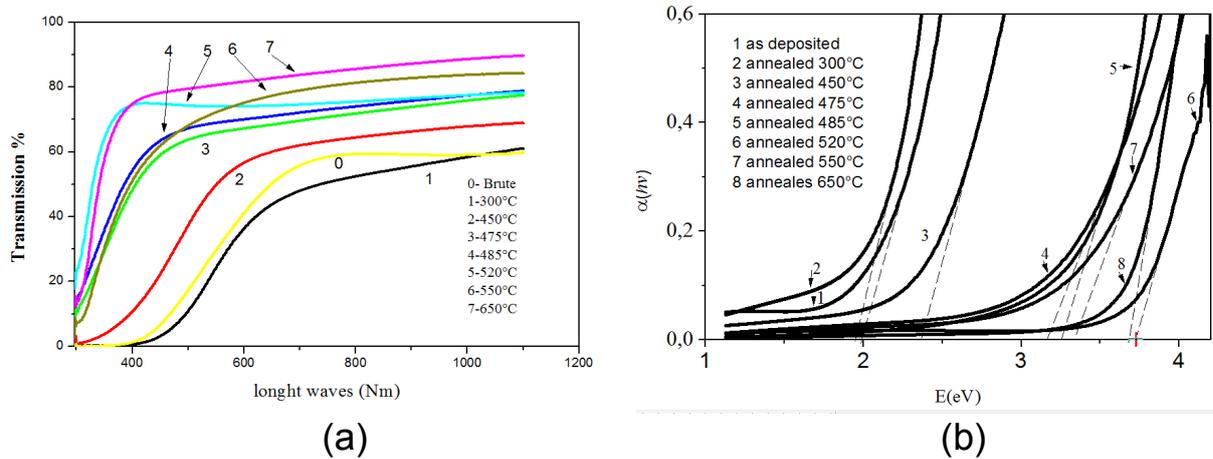

**Figure 13: UV-is spectra of different samples**

In these compounds we consider here the direct transitions occurring during the absorption process [23, 24] and thus the absorption coefficient $\alpha$ writes [27]:

$$\alpha = \frac{C}{h\nu}\cdot[h\nu - Eg]^{(\frac{1}{2})} \quad (3)$$

where C is a constant, h is the Planck constant, h$\nu$ the energy of incident photons, and $E_g$ the direct optical band gap to be determined. As usually in analogous cases, the band gap is determined basing on the formula derived from (3):

$$(\alpha.h\nu)^2 = C^2\cdot[h\nu - Eg] \quad (4)$$

by drawing the $(\alpha.h\nu)^2$ curve versus energy, and extrapolating its linear part to zero. This is done in Figure 12b.

We see that the band gap varies from 1.94 eV for the sample treated at 300 C, presents a maximum at 3.72 eV for the sample treated at 520 C, goes down to 3.33 eV for the sample treated at 550C, and finally takes on the value of 3.68 eV for the sample treated at 650 C. These data hint also at a particular behaviour of the sample treated at 520 C and at an intermediate structure.

Before discussing the band gap variation as a function of the composition parameter x, it is worthwhile to consider the question of the band gap of β-$In_2S_3$. The band gap of this material has been studied in numerous works, and it is interesting to compare our results with results of other studies, that vary as a function of structural details as sample thickness, grain size, defects of various types, and consequently conditions of sample preparation as thermal annealing temperatures, etc.

In β-$In_2S_3$ grown by physical vapour deposition (PVD) the band gap is determined to be 2.1 eV , a value lying in the interval generally reported of  2.0 - 2.4 eV [28, 29]. However in samples synthesized by flash evaporation of β-$In_2S_3$ powders [30], the optical transmission spectra show a slight shift of the absorption edge towards lower wavelengths and the band gap value varies between 2.4 and 3 eV depending on the film thickness and the annealing temperature.

In pure β-$In_2S_3$ there are in fact reports for direct and indirect band gaps ; Sandoval and coworkers measured the values of indirect and direct band gaps and found 1.98 and 2.53 eV, respectively, for samples annealed at 450 C. Such a high value of the band gap may be compared to the value measured in amorphous $In_2S_3$  (2.75 eV) and also in spray pyrolized β-$In_2S_3$ thin films with a In/S ratio of 2/3 in the solution (band gap 2.67 eV) in which a small decrease is observed when annealing (in samples annealed in vacuum at 400C, the gap is 2.62 eV) [31] . Such high values of band gap are reported in various other types of preparations:
- in layers deposited by close-spaced evaporation, the band gap was observed to increase as the effect of the substrate temperature (from 2.09 eV for a substrate temperature of 200C to 2.52 eV for a substrate temperature of 300 C) [32].
- in films prepared by sulfurization of amorphous indium films it is observed also that the band gap value depends on the annealing temperature (from 3 eV to 2.6 eV when the annealing temperature changes from 200 C to 400 C [33]), and on the thickness of the film (from 2.0 to 3.6 eV under variations in the thickness  from 800–450 nm to 50–30 nm) [34]
- in depositions obtained using the sol–gel method a band gap of 2.51 eV is reported [35].
- in films deposited by chemical spray pyrolysis at 300 C, the  band gap decreases slightly from 2.7 to 2.67 eV when the S/In ratio increases from 1.25 to 1.75 [36]

In all these materials, the underlying structure is constituted of crystals of nanometric size and the variation of band gap is determined to be more clearly due to the degree of crystallinity than to the composition (the In/S ratio variation is accompanied only by slight change of the band gap) [37]. Bedir et al reported values of band gap in their compounds, between 2.0 and 2.8 eV [38].

Finally, in nanocrystalline films of indium sulfide obtained by sulfidation of $In_2O_3$ in HS atmosphere, the band gap appears to be 2.0 eV [39], a value close to our determination in the present work, and to other results mentioned above for bulk materials.

The possible mechanisms were discussed in [32], and the authors concluded that *« the grain boundaries might be the dominant source for the presence of structural disorder in these nanocrystalline layers leading to the compensation of the energy band gap »*, predicting that *« annealing of the layers might increase the band gap as the annealing process could decrease the disorder present in the layers »*. According to them, *« the increase of substrate temperature would reduce the density of localized states in the band structure, and also lead to an increase of the band gap »*.

A mechanism that was not considered, that could also influence the value of band gap, is related to the order-disorder phase transitions occurring in $\beta$-$In_2S_3$. This material has indeed a defective spinel structure and presents two structural phase transitions ; indium ions located in tetrahedral interstices disorder at 420 C and those located in octahedral interstices disorder at 780 C, which induces different domain pattern involving domain walls (twins) and translational domains (also called antiphase boundaries) [40]. This could result in an enhancement of band bending around the corresponding walls, the structural disorder generating a local electric field at the origin of localized states in the mid-band gap region, resulting in the decrease in the energy band gap.

What differentiates samples characterized by these different values of band gap remains still undetermined, and systematic studies involving different parameters like crystallinity, composition, size of grains, film thickness, etc, should be performed to better understand the distribution of band gap values in the interval 2 – 3 eV, paying a particular attention to the temperatures of phase transitions at 420 and 780C.

Considering now the question of the solid solution, mention the work by Barreau et al who determined the band gap $E_g$ of indium oxysulfide $\beta$-$In_2S_{3(1-x)}O_{3x}$ grown by physical vapour deposition (PVD) ; they observed that $E_g$ increases monotonically with x in the x-range between 0 and 0.14, from 2.1 eV at x=0 to 2.9 eV at x~ 0.14 [18]. These results are rather consistent with ours described above but we observe that when x continues increasing the $E_g$ evolution is non monotonous. When x tends to 1, which corresponds to the $In_2O_3$ composition, we have a direct band gap around 3.7 eV.

In$_2$O$_3$ has been also the object of numerous studies, with some controversies, but the situation has been clarified by the work of King et al [26] who investigated the electronic structure of this compound. Their main results are the absence of any evidence for an indirect transition, and the theoretical value of direct band gap of 2.93 ± 0.15 eV for the cubic form of this material. Reported values, including our present results, are higher. This was first attributed in most of papers to indirect optical absorption, but an alternative explanation was given thereafter : due to the symmetry of the bixbyite crystal structure (cubic), optical transitions from the valence-band maximum to the conduction band minimum are forbidden, and other transitions from the valence bands to the conduction band are also either forbidden or have only very weak optical transition matrix elements. This mechanism explains the presence of the tail in the optical absorption spectrum below the linear part, as can be seen in Figure 13-b, leading to an overestimated value of E$_g$ of 0.5– 1 eV above the fundamental band gap.

In our case, the measured value of ~3.7 eV is about ~0.7 eV above the predicted value, in good agreement with considerations presented in [39].

## Conclusion

In the present study we showed that it is possible to obtain In$_2$S$_3$ films of good quality by Dr Blade method from a powder synthesized in a chemical bath. The films may be exempt of any crack by adding a small quantity of polyethylene glycol in the preparation, which makes these films potentially interesting for solar cells. When submitted to oxidation, the films remain of good morphological quality, and appear homogeneous.
The oxidation process begins significantly above 420 C, and it appears that at least one intermediate crystal phase takes place in the solid solution, as evidenced by results of thermogravimetric measurements, XRD data, and results of UV-visible spectroscopy. This suggests that the replacement of sulfur atoms by oxygen atoms occurs at different temperatures for the different crystallographic sites, which could lead to several intermediate crystal structures in the phase diagram. Further detailed studies are required to confirm this last assumption and determine the nature of the intermediate phase(s). The present study shows that it is possible to obtain films with a micrometric grain size, of intermediate composition between pure indium sulfide and indium oxide, depending on the annealing temperature in air atmosphere, with a band gap varying continuously between 1.94 (In$_2$S$_3$) and 3.7 eV (In$_2$O$_3$), this last value being overestimated by mechanisms of forbidden transitions between electronic bands.

## Acknowledgements

We are grateful to N. Masquelez for her assistance during TGA experiments, and to D. Cot and B. Rebière as contacts of the electron microscopy service in the European Institute of Membranes (IEM) in Montpellier.